%% file: main.tex
\documentclass[acmtog]{acmart}
\acmSubmissionID{912}

\usepackage[ruled]{algorithm2e} % For algorithms

\usepackage{multirow}
\usepackage{subcaption}
\usepackage{calc}

\SetAlFnt{\small}
\SetAlCapFnt{\small}
\SetAlCapNameFnt{\small}
\SetAlCapHSkip{0pt}

\acmJournal{TOG}

\copyrightyear{2026}
\acmYear{2026}
\setcopyright{cc}
\setcctype{by}
\acmConference[SIGGRAPH Conference Papers '26]{Special Interest Group on Computer Graphics and Interactive Techniques Conference Conference Papers}{July 19--23, 2026}{Los Angeles, CA, USA}
\acmBooktitle{Special Interest Group on Computer Graphics and Interactive Techniques Conference Conference Papers (SIGGRAPH Conference Papers '26), July 19--23, 2026, Los Angeles, CA, USA}
\acmDOI{10.1145/3799902.3811157}
\acmISBN{979-8-4007-2554-8/2026/07}

\begin{document}
% Title portion
\title{Adaptive Interpolation-Synthesis for Motion In-Betweening on Keyframe-Based Animation}

\author{Anton Raël}
\orcid{0009-0000-0710-3166}
\affiliation{%
  \institution{Animaj}
  \city{Paris}
  \country{France}
}
\email{anton@animaj.com}

\author{Julien Boucher}
\orcid{0009-0004-1608-9991}
\affiliation{%
  \institution{Animaj}
  \city{Paris}
  \country{France}
}
\email{julien.boucher@animaj.com}

\author{Antoine Lhermitte}
\orcid{0009-0000-1965-8986}
\affiliation{%
  \institution{Animaj}
  \city{Paris}
  \country{France}
}
\email{antoine@animaj.com}




\begingroup

\begin{teaserfigure}
\includegraphics{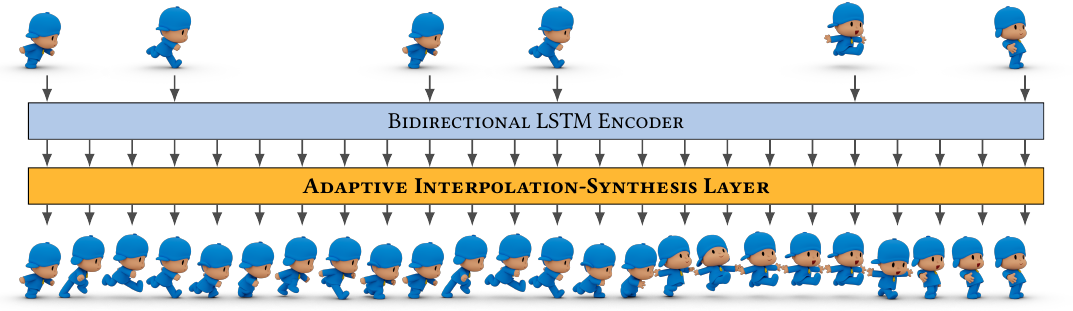}
\captionsetup{skip=0pt}
\caption{Our novel \textbf{Adaptive Interpolation-Synthesis (AIS) layer}, combined with a Bi-LSTM encoder, generates dense 3D animation (bottom) from sparse block poses (top). It produces accurate intermediate poses while preserving motion style, yielding high-quality results that require only minor retakes and accelerate the in-betweening process by up to 3.5×.}
\label{fig:teaser}
\end{teaserfigure}
\endgroup

\input{sections/abstract}

\maketitle

\input{sections/introduction}

\input{sections/related_work}

\input{sections/methodology}

\input{sections/experiments}

\input{sections/limitations_future_work}

\input{sections/conclusion}

\bibliographystyle{ACM-Reference-Format}
\bibliography{bibliography}

\input{figures/curves_detailed}
\input{figures/renderings_sequences}
\input{figures/lafan1_renderings}
\input{figures/block_schedule_impact}
\input{figures/pato_sitting_jumping}

\end{document}

%% file: sections/abstract.tex
\begin{abstract}

Motion in-betweening is one of the most artistically demanding and time-consuming stages of 3D animation, where the expressivity and rhythm of motion are defined. The level of creative control it requires makes it a major production bottleneck, underscoring the need for intelligent tools that assist animators in this process. Although recent deep learning approaches have achieved strong results in motion synthesis and in-betweening, they assume data characteristics, motion styles, and problem formulations that diverge from professional animation workflows. To bridge this gap, we propose a method explicitly aligned with the constraints of motion in-betweening for keyframe-based animation in production environments. At its core, the \emph{Adaptive Interpolation–Synthesis (AIS)} layer mirrors the animator’s creative process by dynamically balancing learned interpolation and direct pose synthesis. In addition, a domain-based input keypose schedule reflects the distribution of production data, improving stylistic consistency and alignment between training and real-world usage. Our method achieves state-of-the-art performance on production data; when integrated into Autodesk Maya, it enables animators to complete in-betweening tasks with a $3.5\times$ speedup.

\end{abstract}

%% file: sections/introduction.tex
\section{Introduction}

In production studios, animators typically begin by manually crafting a sequence of 3D block keyposes derived from a 2D animatic. These sparse, expressive poses act as the structural anchors of the performance and define the storytelling intent and semantics of the motion---they describe \textit{what} the character is doing. Once the block poses are approved, production proceeds to the motion in-betweening phase, where animators connect these poses through coherent intermediate motions. This stage determines \textit{how} the character moves, shaping the flow, emotion, and energy of the animation. As articulated in the seminal \textit{Illusion of Life: Disney Animation}~\cite{thomas1981illusion}, motion in animation follows a set of intertwined principles: anticipation before major actions (Anticipation, Principle~2), gradual acceleration and deceleration (Slow In and Slow Out, Principle~6), deformation conveying weight and flexibility (Squash and Stretch, Principle~1), offset rhythms across body parts (Follow Through and Overlapping Action, Principle~5), control of speed and tempo (Timing, Principle~9), and stylized exaggeration beyond realism (Exaggeration, Principle~10). These principles make motion in-betweening a deeply artistic task that goes beyond simple mathematical interpolation. In practice, this requires animators to create additional keyposes called \textit{breakdown poses} and manually refine interpolation curves in the graph editor by manipulating tangents and timing. This process is labor-intensive and has become a bottleneck in modern production pipelines. At the same time, the growth of digital platforms---with shorter turnaround cycles and higher content demand---has intensified the need for more efficient workflows. Developing intelligent tools to assist animators during in-betweening has thus become an essential goal.

Deep learning offers a data-driven alternative: neural networks can learn from previous episodes featuring a character and capture the stylistic choices made by animators during the in-betweening phase. While recent deep learning methods have shown promising results, they follow paradigms that diverge from production workflows in three main ways. \textbf{Nature of the data:} Prior methods rely on motion capture (mocap) datasets composed of dense, frame-wise pose recordings. Consequently, most models treat motion as a high-dimensional signal, regressing full poses frame-by-frame. In contrast, keyframe-based animation relies on parametric interpolation curves that reduce the effective dimensionality of the motion. \textbf{Motion style:} Developed primarily to synthesize physically plausible human motion, many methods assume smoothness and continuity in their modeling. In contrast, animation is not constrained to realistic human movement and is often intentionally stylized---featuring exaggerated timing, held poses, and snappy transitions. \textbf{Problem formulation:} Most existing work treats motion in-betweening as a completion or inpainting task with large, random temporal gaps, requiring the model to infer both \textit{what} happens and \textit{how}. In production, however, block keyposes are semantically meaningful poses that already define \textit{what} occurs. The animator’s task during in-betweening is to generate short, stylistically coherent transitions that express \textit{how} the motion unfolds.

In this work, we propose a method tailored to the specific challenges of keyframe-based animation in production workflows, with two primary contributions. First, we introduce an \emph{Adaptive Interpolation–Synthesis (AIS)} Layer, which mimics an animator’s workflow by dynamically blending direct pose synthesis with explicit interpolation. Second, we study the impact of input keypose schedules during training, demonstrating that selecting candidate input keyposes according to a domain-based schedule produces more stylistically coherent and consistent motion than random schedules. Our model, combining a bidirectional LSTM with the AIS layer (Fig.~\ref{fig:teaser}), achieves state-of-the-art results on production data. We evaluate our approach using a custom metric designed to strongly correlate with animator retakes time, and quantify time savings in a real production setting where our tool is integrated into Autodesk Maya.

%% file: sections/related_work.tex
\section{Related Work}

\paragraph{Architectures for Motion In-Betweening}

Early approaches to motion in-betweening successfully used recurrent networks like RNNs and LSTMs \cite{martinez2017rnn, harvey2018rtn, harvey2020robust} to handle short-term gaps. As research shifted toward challenging long-term gaps of 30–40 frames, the need for long-range dependency modeling and generative synthesis led to the adoption of Transformers \cite{duan2021singleshot, oreshkin2022delta, sridhar2022tweentransformers, qin2022twostage, kim2022conditionalmib, mo2023citl, studer2024factorized, pinyoanuntapong2024mmmgenerativemaskedmotion, seokhyeon2024keyframeprediction, akhoundi2025silk, goel2025generativemotioninfillingimprecisely} and diffusion models \cite{cohan2024condmi, peng2025autoregressive}. In contrast, keyframe-based animation in production involves short intervals (5–10 frames on average) where motion semantics are already captured by the input block poses. To align with this specific problem formulation, we use a bidirectional LSTM encoder \cite{schuster1997birnn, hochreiter1997lstm, graves2005bilstm} whose bidirectional local inductive bias is better suited for short-term gaps.

\paragraph{Prediction Head Formulation.}

Motion inpainting methods such as CondMDI~\cite{cohan2024condmi} predict poses via frame-by-frame absolute regression in the pose space. This representation is effective for predicting complex breakdown poses. However, in keyframe-based animation, most parts of the motion are defined by low-parametric interpolation curves---step, linear, or spline---with spatial correlation between body parts, reducing the effective dimensionality of the motion. In such cases, this high-dimensional, unconstrained head prediction leads to temporal instability and jitter. Expressing motions in a latent motion manifold, as in CITL~\cite{mo2023citl}, can reduce the effective dimensionality of the prediction. However, this reduction leads to spatial and temporal information loss: spatially, it compromises pose reconstruction fidelity when projecting the latent motion into the pose space, and temporally, it over-smoothes complex, high-frequency motions. \citet{studer2024factorized} propose a Bézier Motion Model that generates Bézier curves between sparse constraints. Similarly, \citet{oreshkin2022delta} explicitly interpolate in the full pose space by using Spherical Linear Interpolation and predicting the residual offsets. While reducing the effective dimensionality of the prediction and guaranteeing fidelity to the input keyposes, both methods rely on a fundamental smoothness assumption in their interpolation functions. Consequently, they are not suited to model snappy and expressive motion styles common in production animation. Building on these insights, we propose a novel prediction head formulation: the Adaptive Interpolation–Synthesis (AIS) layer. Following \citet{studer2024factorized} and \citet{oreshkin2022delta}, we explicitly interpolate between input keyposes, constraining the search space while guaranteeing pose reconstruction fidelity. However, instead of relying on smooth, fixed interpolation functions, we predict the interpolation weights to adapt to any motion style. We supplement this interpolation path with a direct synthesis path---inspired by CondMDI~\cite{cohan2024condmi}---to capture complex breakdown poses. Finally, instead of using an offset formulation, we blend the two paths through a gating mechanism that dynamically adjusts the effective dimensionality of the prediction to the complexity of the motion.

\paragraph{Masking Strategy in Inpainting}

The design of masking strategies is a significant research area in the inpainting literature, where studies have repeatedly shown the benefits of domain-based masks that imitate real-world data distributions, such as object-based masks for image inpainting \cite{shimosato2024maskoptimization,zheng2022cmganimageinpaintingcascaded}, strategies simulating real-world masks for time-series imputation \cite{qian2024imputationmaskingstrategy}, and clinically-informed masks for medical data inpainting \cite{qian2025randommissingnessclinicallyrethinking}. In motion inpainting, however, existing methods still adopt generic approaches such as random sampling \cite{mo2023citl,cohan2024condmi} or uniform spacing \cite{oreshkin2022delta}. In production animation, by contrast, input poses are not random but correspond to block keyposes that encode the semantic content of the motion. Moreover, for expressive motion styles with held poses and snappy transitions, random schedules during training introduce ambiguity in the timing of transitions, often resulting in oversmoothed motion. Following the findings in other inpainting domains, we address this gap by proposing a domain-based schedule that approximates production block poses using the keypose extraction algorithm of~\cite{miura2014adaptivekeyposeextraction}.

%% file: sections/methodology.tex
\input{figures/schema_architecture}

\section{Method}

\input{sections/methodology/data_representation}
\input{sections/methodology/architecture}

\input{sections/methodology/input_keyframes}

\input{sections/methodology/loss}

%% file: figures/schema_architecture.tex
\begin{figure*}[t]
\centering
\includegraphics[width=\linewidth]{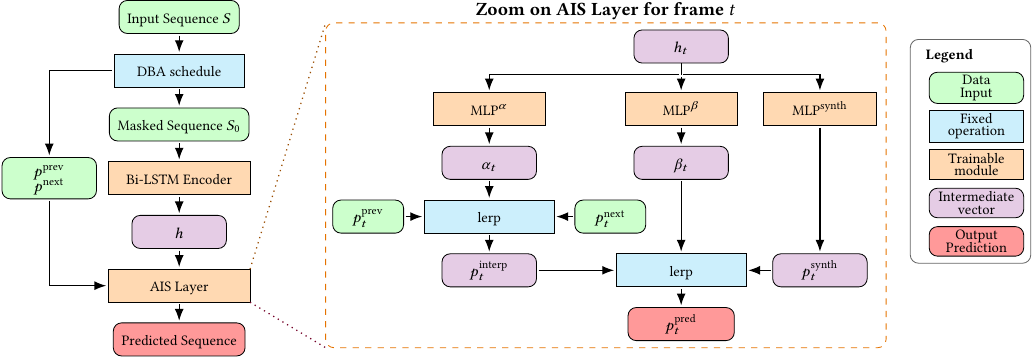}
\caption{Overview of our AIS-BiLSTM architecture.
\textbf{Left:} The Domain-Based Algorithm (DBA) schedule selects training input keyposes to form the masked sequence $S_0$, encoded by a Bi-LSTM into hidden states $h$ for the AIS layer.
\textbf{Middle:} AIS layer zoom at frame $t$. Hidden state $h_t$ drives three MLPs predicting interpolation weights $\alpha_t$, synthesis pose $p_t^{\text{synth}}$, and blending weights $\beta_t$. $\alpha_t$ defines the linear interpolation (lerp) of $p_t^{\text{prev}}$ and $p_t^{\text{next}}$ into $p_t^{\text{interp}}$, while $\beta_t$ adaptively blends $p_t^{\text{interp}}$ and $p_t^{\text{synth}}$ for the final prediction $p_t^{\text{pred}}$.
\textbf{Right:} Legend for the diagram elements.}
\label{fig:architecture}
\end{figure*}

%% file: sections/methodology/data_representation.tex
\subsection{Data representation and notations}

A character's 3D pose is defined by a set of rig controllers' attribute values. Controllers include continuous (translations, rotations, scales) and discrete (FK/IK switches, facial selectors) parameters. Our method predicts only the continuous parameters; discrete values are simply held constant from the last input keypose. Rotations use the 6D representation \cite{zhou2020continuityrotationrepresentationsneural} to avoid gimbal lock and ensure continuity.  Translations and scales are normalized by three times their training set standard deviation to map values approximately to $[-1,1]$. Thus, a full character pose is represented as a single vector $p \in \mathbb{R}^D$, where $D$ is the total pose dimension.

The motion in-betweening task consists of generating a dense animation sequence $\mathcal{S} = \{p_0, p_1, \ldots, p_{N-1}\}$ of $N$ pose vectors, given only a sparse set of $K$ input poses at indices $\mathcal{I} = \{k_0, k_1, \ldots, k_{K-1}\}$, where $k_0 = 0$ and $k_{K-1} = N-1$. The network input $S_0 \in \mathbb{R}^{N\times(D+1)}$ is constructed by zeroing out non-input poses and appending a binary mask:

\begin{equation}
S_0[t] =
\begin{cases}
[p_t, 0] & \text{if } t \in \mathcal{I}, \\
[0_D, 1] & \text{otherwise}.
\end{cases}
\end{equation}

For any frame $t$, we define its closest surrounding input keyposes: $p_t^{\text{prev}} = p_{l(t)}$ and $p_t^{\text{next}} = p_{r(t)}$ with $l(t) = \max \{ k \in \mathcal{I} \mid k \le t \}$ and $r(t) = \min \{ k \in \mathcal{I} \mid k \ge t \}$. An illustration of $p_t^{\text{prev}}$ and $p_t^{\text{next}}$ is provided in Fig.~\ref{fig:ais-schema}(2).

%% file: sections/methodology/architecture.tex
\subsection{AIS-BiLSTM Architecture}

Our AIS-BiLSTM architecture (Fig.~\ref{fig:architecture}) consists of a bidirectional LSTM encoder followed by an Adaptive Interpolation-Synthesis (AIS) prediction head, which constitutes our main architectural contribution. The entire system—comprising the Bi-LSTM and AIS components—is trained jointly in an end-to-end manner.

\subsubsection{Bidirectional LSTM Encoder}
A multi-layer Bi-LSTM encodes the masked input sequence $S_0$, processing it in both directions to capture temporal context from past and future input keyposes. The output is a sequence of hidden states
\begin{equation}
h = \text{BiLSTM}(S_0)  \in \mathbb{R}^{N\times H}.
\end{equation}

\input{figures/illustration_schema_ais}

\subsubsection{Adaptive Interpolation-Synthesis (AIS) Layer}

The design of the AIS layer is inspired by the two workflows animators employ when creating in-between poses:

\begin{itemize}
    \item \textbf{Explicit Interpolation:} For many transitions, animators explicitly define the motion between two keyposes as an interpolation function. This can involve selecting a predefined interpolation type (e.g., linear, spline), directly manipulating curves in a graph editor by adjusting the tangents, or using tools like a Tween Machine to set a precise interpolation coefficient.
    \item \textbf{Pose Synthesis:} For more complex or expressive actions, such as creating a critical breakdown pose or an exaggerated effect like an overshoot, a simple interpolation is insufficient. In these cases, animators directly manipulate the character's rig to sculpt a new pose, effectively synthesizing a value that may lie outside the interpolation space defined by the bounding keyposes.
\end{itemize}

Our AIS layer explicitly models this decision-making process. It formulates the final pose prediction as an adaptive blend of two distinct paths, at the frame and controller levels, as illustrated in Fig.~\ref{fig:ais-schema}. Mathematically, for each frame $t$, the hidden state $h_t$ is used to compute two paths and a gate coefficient:

\paragraph{Interpolation Path} This path mimics the first workflow and learns to predict an explicit interpolation between the previous and next input keyposes (Fig.~\ref{fig:ais-schema}(3)). Interpolation is performed directly in pose space: translations and scales in their respective 3D vector spaces, and rotations in the continuous 6D representation. A frame-specific interpolation weight vector $\alpha_t \in [0, 1]^D$ is predicted from the hidden state $h_t$ using a small Multi-Layer Perceptron (MLP) with a final sigmoid activation. The resulting interpolated pose is computed using an element-wise product (Hadamard product), denoted by $\odot$:
\begin{align}
    \alpha_t &= \sigma(\text{MLP}^{\alpha}(h_t)), \label{eq:alpha}\\
    p_t^{\text{interp}} &= (1 - \alpha_t) \odot p_t^{\text{prev}} + \alpha_t \odot p_t^{\text{next}}. \label{eq:interp}
\end{align}

\paragraph{Synthesis Path} This path emulates the second workflow, directly regressing a new pose from the temporal context (Fig.~\ref{fig:ais-schema}(4)). This enables the model to generate poses that differ significantly from the previous and next input keyposes, or lie just outside their range. The pose $p_t^{\text{synth}}$ is predicted from the hidden state $h_t$ using a small Multi-Layer Perceptron (MLP).
\begin{equation}
    p_t^{\text{synth}} = \text{MLP}^{\text{synth}}(h_t).
    \label{eq:synth}
\end{equation}

\paragraph{Blending Gate} This learned gating mechanism determines the appropriate contribution from each path, allowing the model to adapt its strategy on a frame-by-frame basis for each pose dimension (Fig.~\ref{fig:ais-schema}(5)). The blending coefficient vector, $\beta_t \in [0, 1]^D$, is predicted by another MLP. The final predicted pose, $p_t^{\text{pred}}$, is then calculated as a linear interpolation between the two paths' outputs, gated by $\beta_t$:
\begin{align}
    \beta_t &= \sigma(\text{MLP}^{\beta}(h_t)), \label{eq:beta}\\
    p_t^{\text{pred}} &= (1 - \beta_t) \odot p_t^{\text{interp}} + \beta_t \odot p_t^{\text{synth}}. \label{eq:ais-blend}
\end{align}

This mechanism allows per-attribute adaptive blending between interpolation and synthesis, mirroring the animator’s workflow.

%% file: figures/illustration_schema_ais.tex
\begin{figure*}[t]
  \centering
  \includegraphics[width=\linewidth,trim=2mm 1mm 2mm 1mm,clip]{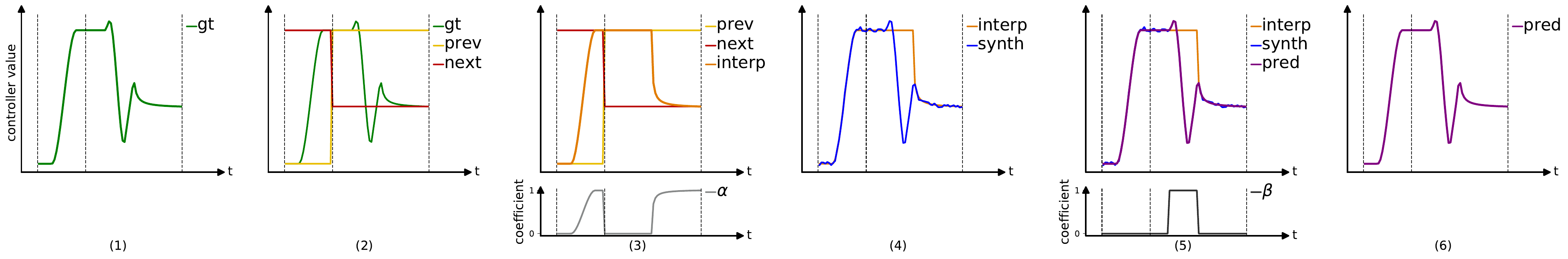}
\caption{Step-by-step visualization of the Adaptive Interpolation-Synthesis (AIS) layer's operation on a single controller value over time.
Vertical dotted lines indicate input keyposes.
\textbf{(1)} \textbf{Ground-truth (GT)}: stylized motion with held poses, snappy transitions, and complex anticipation and overshoot effects.
\textbf{(2)} \textbf{Input Keyposes}: step functions of the previous (\texttt{prev}) and next (\texttt{next}) input keypose values.
\textbf{(3)} \textbf{Interpolation Path}: predicted time-varying coefficient $\alpha$ to blend the \texttt{prev} and \texttt{next} into an \texttt{interp} curve.
\textbf{(4)} \textbf{Synthesis Path}: directly regressed \texttt{synth} motion capturing complex details, yet exhibiting jitter on simpler segments.
\textbf{(5)} \textbf{Blending Gate}: predicted coefficient $\beta$ to adaptively weight the \texttt{interp} and \texttt{synth} signals based on local motion complexity.
\textbf{(6)} \textbf{Final Prediction}: resulting \texttt{pred} curve combining interpolation stability with synthesis expressiveness to reconstruct the target motion.
}
\label{fig:ais-schema}
\end{figure*}

% \begin{figure*}[t]
%   \centering
%   % Prefer a PDF exported by your script: save_path="figs/plots_schema_paper.pdf"
%   \includegraphics[width=\linewidth,trim=2mm 1mm 2mm 1mm,clip]{figures/illustration_schema_ais/plots_schema_paper.pdf}
% \caption{Step-by-step visualization of the Adaptive Interpolation-Synthesis (AIS) layer's operation on a single controller value over time.
% Vertical dotted lines indicate input keyposes.
% \textbf{(1)} \textbf{Ground-truth (GT)}:complex, stylized motion.
% \textbf{(2)} \textbf{Input Keyposes}, where the constant values of the previous (\texttt{prev}) and next (\texttt{next}) input keyposes are represented as step functions across the interpolation interval.
% \textbf{(3)} \textbf{Interpolation Path}, where the model predicts a time-varying coefficient $\alpha$ to blend the \texttt{prev} and \texttt{next} signals, producing a baseline \texttt{interp} curve.
% \textbf{(4)} The \textbf{Synthesis Path}, where the model directly regresses the \texttt{synth} signal. This path can generate complex motion but may exhibit noise on simpler segments.
% \textbf{(5)} The \textbf{Blending Gate}, where a predicted coefficient $\beta$ determines the blend between the \texttt{interp} and \texttt{synth} paths, allowing the output to adapt to the motion's complexity.
% \textbf{(6)} The \textbf{Final Prediction}, where the \texttt{pred} curve successfully combines the stability of the interpolation path with the expressiveness of the synthesis path to accurately reconstruct the ground-truth motion.}
% \label{fig:ais-schema}
% \end{figure*}

%% file: sections/methodology/input_keyframes.tex
\subsection{Input Keypose Selection}

\subsubsection{Domain-Based Algorithm (DBA)}

In production settings, input block keyposes are not randomly distributed; they correspond to specific poses that encode the semantic intent of the motion and follow what we term a \textbf{production block schedule}. However, this block schedule is not stored in our data, as the final animation files contain a mixture of block keyposes and additional keyposes introduced during the in-betweening phase that are not distinguishable.

To address this limitation, we adopt the keypose extraction method proposed by \cite{miura2014adaptivekeyposeextraction} to identify the set of input keyposes. Originally developed for motion capture data, this algorithm processes a dense sequence of poses and extracts keyposes corresponding to local minima in motion speed at multiple frequency scales. Since our motion style is characterized by snappy transitions and held poses, we hypothesize that this approach serves as an effective proxy for production block schedules. We refer to this keypose selection method as the \textbf{Domain-Based Algorithm input keypose schedule} (or simply \textbf{DBA schedule}).

\subsubsection{Input Keypose Schedule Augmentation}
\label{sec:input_keypose_schedule_augmentation}

Since the keyposes selected by the Domain-Based Algorithm are only a proxy for true production block keyposes, we apply random augmentation to improve the generalization capabilities of the model. We augment the DBA input keypose schedule by randomly adding or removing keyposes, thereby varying the density of the schedule. To model differences in timing choices---such as cycle offsets or minor placement preferences---we also apply random shifts to keypose indices, both globally (shifting the entire input keypose schedule) and locally (jittering individual frames). We define three augmentation levels---see Supplementary Material for more details---with progressively higher probabilities and magnitudes of add/remove/shift operations, which allow us to test the impact of augmentation strength during training.

%% file: sections/methodology/loss.tex
\subsection{Loss Function}

We train the model using a multidimensional L1 reconstruction loss. To prevent controllers with larger error magnitudes from dominating the gradient, we employ an automatic loss-weighting strategy inspired by multi-task learning \cite{liebel2018multitask}. The loss $\mathcal{L}$ for frame $t$ is:

\begin{equation}\mathcal{L} = \sum_{d=0}^{D-1} \left( \frac{1}{2\sigma_d^2} | p_{t,d}^{\text{pred}} - p_{t,d}^{\text{gt}} |_1 + \log(1 + \sigma_d^2) \right)
\end{equation}

where $\sigma_d$ is a learnable parameter that scales the L1 error for each pose dimension $d$. The logarithmic term regularizes $\sigma_d$, allowing the network to adaptively balance the contribution of various controller attributes during training.

%% file: sections/experiments.tex
\input{figures/ablation_study}
\section{Experiments and results}

\input{sections/experiments/datasets}

\input{sections/experiments/metric}

\input{sections/experiments/ablation_study_architectural}

\input{sections/experiments/ablation_study_training_input_keyframes}

\input{figures/sota_comparison_table}
\input{sections/experiments/comparison_sota}

\input{sections/experiments/production_workflow_evaluation}

%% file: figures/ablation_study.tex
\begin{table*}[t]
\setlength{\tabcolsep}{2pt}
\centering
% \caption{Ablation studies of different models on three evaluation sets: the Algorithmic Test Set (Algo. Set) with DBA input keypose schedule, the Random Test Set (Rand. Set) with random masking at $r=0.9$, and the Production Test Set (Prod. Set) with production block schedules. \textbf{Left:} Prediction head variants. Our AIS layer consistently outperforms alternative heads across all sets. \textbf{Right:} Training input keypose schedules. Aligning the training and inference input keypose schedules improves performance; on the production set, the DBA input keypose schedule with level-2 augmentation achieves the best results.}
% \caption{Ablation studies on three evaluation sets: Algorithmic (Algo.), Random (Rand.), and Production (Prod.). \textbf{Left:} Prediction head variants. Our AIS layer consistently outperforms alternative heads across all sets. \textbf{Right:} Training input keypose schedules. Aligning the training and inference input keypose schedules improves performance; on the production set, the DBA input keypose schedule with level-2 augmentation achieves the best results.}
\caption{
    \textbf{Ablation studies on three evaluation sets:} Algorithmic (\textbf{Algo.}), Random (\textbf{Rand.}), and Production (\textbf{Prod.}).
    \textbf{Left:} Comparison of prediction head variants shows that our \textbf{AIS layer} consistently outperforms alternative prediction heads.
    \textbf{Right:} Comparison of training input schedules highlights the importance of aligning training and inference schedules. On the Prod. Set, the DBA schedule with second level of augmentation outperforms random schedules.
}
% \caption{Ablation studies. We evaluate different model on the Algorithmic Test Set (Algo. Set) containing DBA block schedule, the Random Test Set (Rand. Set) with a random schedule of masked ratio r=0.9, and the Production Test Set (Prod. Set) with a real production schedule. \textbf{Left:} Prediction head variants. Our AIS Layer outperforms other head prediction on all test sets. Right: Input keyframes schedule. Aligning the training schedule with the inference schedule yields the best results. On the production set, our DBA schedule with the second level of augmentation yiels the best result, confirming our design choice }
\label{tab:ablation_combined}
\begin{minipage}[t]{0.48\textwidth}
\centering
\textbf{(a) Prediction head variants} \\
\begin{tabular}{@{}lcccccc@{}}
\toprule
\multirow{2}{*}{Prediction head} & \multicolumn{2}{c}{Algo. Set} & \multicolumn{2}{c}{Rand. Set} & \multicolumn{2}{c}{Prod. Set} \\
\cmidrule(lr){2-3} \cmidrule(lr){4-5} \cmidrule(lr){6-7}
& STL1 $\downarrow$ & NPSS $\downarrow$ & STL1 $\downarrow$ & NPSS $\downarrow$ & STL1 $\downarrow$ & NPSS $\downarrow$ \\
\midrule
Direct synthesis         & 13.50 & 1.69          & 14.97 & 2.07          & 32.70 & 3.30 \\
SLERP+offset               & 5.12  & 1.18          & 6.15  & 1.22          & 4.15  & 0.62 \\
Linear6D+offset            & 4.54  & 1.10          & 5.58  & 1.13          & 3.20  & 0.54 \\
Previous+offset            & 2.11  & 0.73          & 4.94  & 1.34          & 2.13  & 0.50 \\
Interp.-only        & 1.84  & 0.61          & 4.16  & 1.14          & 1.86  & 0.43 \\
Interp.+offset     & 1.72  & 0.55          & 4.19  & 1.13          & 1.87  & \textbf{0.39} \\
Fixed gate & 7.90  & 0.55          & 9.87  & 1.05          & 17.66 & 0.91 \\
% Interp.-offset      & \textbf{1.46}  & \textbf{0.37} & \textbf{3.82}  & \textbf{0.94} & 1.86  & 0.43 \\
\textbf{AIS}                 & \textbf{1.53}  & \textbf{0.46}          & \textbf{3.87}  & \textbf{1.05}          & \textbf{1.77}  & 0.40 \\
\bottomrule
\end{tabular}
\end{minipage}
\hfill
\begin{minipage}[t]{0.51\textwidth}
\centering
\textbf{(b) Training input keypose schedule} \\
\begin{tabular}{@{}lcccccc@{}}
\toprule
\multirow{2}{*}{\shortstack{Input keypose \\ schedule}} & \multicolumn{2}{c}{Algo. Set} & \multicolumn{2}{c}{Rand. Set} & \multicolumn{2}{c}{Prod. Set} \\
\cmidrule(lr){2-3} \cmidrule(lr){4-5} \cmidrule(lr){6-7}
& STL1 $\downarrow$ & NPSS $\downarrow$ & STL1 $\downarrow$ & NPSS $\downarrow$ & STL1 $\downarrow$ & NPSS $\downarrow$ \\
\midrule
DBA              & 1.41 & \textbf{0.35} & 4.71 & 1.05 & 2.07 & 0.48\\
DBA + aug. lvl 1 & \textbf{1.40} & 0.45 & 4.15 & 1.1 & 2.02 & 0.46 \\
\textbf{DBA + aug. lvl 2} & 1.53 & 0.46 & 3.87 & 1.05 & \textbf{1.77} & \textbf{0.40} \\
DBA + aug. lvl 3 & 1.74 & 0.53 & 3.72 & 0.98 & 1.81 & 0.42 \\
% Random (r=0.5)   & 2.40 & 0.67 & 3.50 & 0.82 & 2.13 & 0.48 \\
% Random (r=0.6)   & 2.26 & 0.63 & 3.17 & 0.75 & 2.03 & 0.45 \\
% Random (r=0.7)   & 2.42 & 0.68 & 3.00 & 0.74 & 2.13 & 0.44 \\
Random (r=0.8)   & 2.57 & 0.70 & 3.01 & 0.79 & 1.99 & 0.46 \\
Random (r=0.9)   & 2.36 & 0.66 & \textbf{2.83} & \textbf{0.73} & 2.23 & 0.45 \\
Random (r=0.95)  & 2.90 & 0.85 & 3.53 & 0.85 & 2.30 & 0.45 \\

\bottomrule
\end{tabular}
\end{minipage}
\end{table*}

%% file: sections/experiments/datasets.tex
\subsection{Datasets and Evaluation Setup}

Our primary evaluation is conducted on a proprietary \textbf{In-House Dataset} of professionally handcrafted animations, comprising 8.5 hours of motion at 24 fps for a single character. For training, scenes are randomly trimmed to a fixed length of 224 frames to ensure compatibility with the baseline~\cite{cohan2024condmi}, although our model can handle variable-length sequences. After processing, each pose is represented by a $D=596$-dimensional vector. We normalize the root controller, which represents the character's absolute position, to be relative to the first frame of each sequence.

The dataset is partitioned into a 95\% training set and a 5\% \textbf{Held-Out Set} for testing. We evaluate all models on three datasets with different inference input keypose schedules. (1) The \textbf{Algorithmic Test Set} is the Held-Out Set with a DBA inference schedule, measuring performance on our algorithmic schedule. (2) The \textbf{Random Test Set} is the \textbf{Held-Out Set} following a random input keypose schedule with a ratio $r=0.9$ of masked frames, enabling fair comparison with SOTA models trained on random input keypose schedules. (3) The \textbf{Production Test Set} consists of 5 minutes of production data with ground-truth block keyposes, evaluating the generalization to real production block schedules.

%% file: sections/experiments/metric.tex
\subsection{Metrics}

\paragraph{Shift-Tolerant L1 (STL1) Distance}
Standard L1 metrics are highly sensitive to small temporal misalignments. In production, such offsets are trivial to fix by shifting poses, so high error may not reflect long animator retakes. We therefore define the \textbf{Shift-Tolerant L1 (STL1) distance}, which isolates pose prediction errors from timing offsets. Given keyframes $I = \{k_0, \dots, k_{K-1}\}$, each ground-truth subsequence between $k_i$ and $k_{i+1}-1$ has length $L_i$. For each, we search over the predicted sequence for the best alignment within a shift window $\delta \in [-L_i, L_i]$. The error is the minimum L1 distance over this window, summed across all subsequences and normalized by the total number of frames $N$:
\begin{equation}
\text{STL1} = \frac{1}{N} \sum_{i=0}^{K-2} \min_{\delta \in [-L_i, L_i]} \sum_{j=0}^{L_i-1} \big\| p^{\text{gt}}_{k_i+j} - p^{\text{pred}}_{k_i+\delta+j} \big\|_1.
\label{eq:stl_definition}
\end{equation}

\paragraph{NPSS}
We also report the Normalized Power Spectrum Similarity (NPSS) \cite{gopalakrishnan2018npss}, a standard metric for in-betweening. NPSS compares the frequency distributions of predicted and ground-truth joint trajectories, capturing global temporal characteristics such as rhythm and style. It complements STL1 by focusing on long-term stylistic fidelity rather than frame-level accuracy.

%% file: sections/experiments/ablation_study_architectural.tex
\subsection{Ablation Study on AIS Prediction Head}

We validate our primary contribution, the \textbf{Adaptive Interpolation–Synthesis (AIS) Layer}, by comparing it with different prediction head variants. All variants are trained jointly with a Bi-LSTM encoder, following the DBA schedule with the second level of augmentation. First, we construct a \textbf{Direct Synthesis} prediction head that regresses the pose directly from the hidden state ($p_t^{\text{pred}} = p_t^{\text{synth}}$), representing the simplest non-interpolative approach. Second, we implement a group of baselines that follow the “offset” strategy of~\citet{oreshkin2022delta}, where a learned offset is added to a fixed interpolation ($p_t^{\text{pred}} = \operatorname{fixed\_interp}(...) + \text{MLP}(h_t)$). We test three interpolation functions: \textbf{Previous} (last keypose), \textbf{Linear6D} (linear interpolation of rotations in 6D representation, with translations and scales interpolated linearly), and \textbf{SLERP} (spherical linear interpolation of rotations in quaternion space, with translations and scales interpolated linearly). Finally, we perform ablations of our AIS design with three variants: an \textbf{Interpolation-only} head ($p_t^{\text{pred}} = p_t^{\text{interp}}$) to assess the necessity of synthesis, an \textbf{Interpolation+Offset} head ($p_t^{\text{pred}} = p_t^{\text{interp}} + p_t^{\text{synth}}$) that removes adaptive gating and replaces it with a simple additive combination, and a \textbf{Fixed Gate} head ($p_t^{\text{pred}} = 0.5 \cdot p_t^{\text{interp}} + 0.5 \cdot p_t^{\text{synth}}$) that uses a constant blending coefficient instead of learning it, to test the importance of the learned gating mechanism. We compare all variants both quantitatively and qualitatively (Table~\ref{tab:ablation_combined}(a), Fig.~\ref{fig:controllers:big}).

\textbf{Direct synthesis.} Quantitatively, this baseline performs poorly across all datasets and metrics. Qualitatively (Fig.~\ref{fig:ctrl:lastlayer:direct}), it can generate correct breakdown poses for complex motion cycles, but for simpler motions its outputs contain jitter and shift from the input keyposes. This reflects the difficulty of regressing full poses directly and justifies the need for an explicit interpolation term.

\textbf{Fixed Interpolation with Offset.} Quantitatively, models that predict an offset with a fixed interpolation---i.e. \textbf{SLERP+Offset}, \textbf{Linear6D+Offset}, \textbf{Previous+Offset}---outperform \textbf{Direct Synthesis} but consistently underperform models with a learned interpolation module (\textbf{Interpolation-only}, \textbf{Interpolation+Offset}, \textbf{AIS}). Qualitatively (Fig.~\ref{fig:ctrl:lastlayer:mathinterp}), they are more stable than \textbf{Direct Synthesis} but are too strongly biased by their underlying interpolation function: they often collapse toward near-zero offsets, producing overly smooth motions in the case of \textbf{SLERP+Offset} and \textbf{Linear6D+Offset}, or defaulting to a "last-keypose copy" behavior for \textbf{Previous+Offset}. This underscores the limitations of fixed interpolation formulas and highlights the advantage of a learned interpolation module that does not inherit these biases and can adapt to any motion style.

\textbf{AIS Ablations.} Quantitatively, the \textbf{AIS} layer clearly outperforms the \textbf{Interpolation-only}, \textbf{Interpolation+Offset}, and \textbf{Fixed Gate} variants across all test sets. Qualitatively (Fig.~\ref{fig:ctrl:lastlayer:learnedinterp}), the  \textbf{Fixed Gate} prediction head shifts from the input keypose values, as the direct synthesis variant.  The \textbf{Interpolation-only} and \textbf{Interpolation+Offset} models struggle to generate accurate breakdown keyposes in typical motion cycles or to produce anticipation effects, whereas the \textbf{AIS} layer successfully synthesizes these poses. This demonstrates the necessity of a direct synthesis module for complex motions where intermediate poses differ from the input keyposes, and highlights the superiority of the learned gating mechanism over both a simple offset formulation and a fixed blending coefficient.

%% file: sections/experiments/ablation_study_training_input_keyframes.tex
\subsection{Ablation Study on Training Input Keypose Schedules}

This section evaluates our second methodological contribution: the use of a Domain-Based Algorithm (DBA) to select input keyposes during training. We compare our DBA input keypose schedule---with different augmentation levels described in Section~\ref{sec:input_keypose_schedule_augmentation}---against random input keypose schedules with various masking ratios $r$ (e.g., $r=0.8$ means 80\% of frames are randomly masked).
Performance, reported in Table~\ref{tab:ablation_combined}~(b), is measured across the three test sets, each following different inference input keypose schedules.

\paragraph{Effectiveness of the DBA Schedule compared to Random Schedules.}

Quantitative results on the Algo. Set---i.e., the test set following the inference DBA schedule---show that training with the same DBA schedule yields the best performance. As the amount of random augmentation increases, performance progressively degrades, with purely random input keypose schedules performing the worst. This confirms our hypothesis that domain-based input keypose schedules outperform random ones on this domain, consistent with findings in other inpainting tasks. Qualitative results in Fig.~\ref{fig:ctrl:blocksched:random} reveal two issues with random training schedules. First, they tend to produce overly smooth, averaged transitions (Fig.~\ref{fig:ctrl:blocksched:random}, left), whereas the ground-truth style typically consists of two held poses connected by a sharp, snappy transition. When input keyposes are sampled randomly, the model cannot infer when such transitions occur, making the in-betweening task a one-to-many problem. Under L1 loss optimization, the model converges toward the temporal average of all possible transitions, resulting in an oversmoothed motion. Our DBA algorithm removes this ambiguity by deterministically selecting consistent keyposes, allowing the model to learn sharper, more temporally coherent predictions. Second, random schedules produce noisy outputs with undesirable additional motions between input keyposes (Fig.~\ref{fig:ctrl:blocksched:random}, right). Random schedules may omit semantically meaningful keyposes during training, leading the model to infer missing motion content. By selecting consistently these semantically meaningful keyposes, the DBA schedule allows the model to focus only on connecting these poses through the correct intermediate poses and style.

\paragraph{Augmenting the DBA for better generalization.}

Results in Table~\ref{tab:ablation_combined}~(b) and in Fig.~\ref{fig:ctrl:blocksched:aug} for the Rand. and Prod. Sets show that the model trained solely on the deterministic DBA input keypose schedule struggles to generalize to other scheduling patterns, while augmenting the DBA schedule with random transformations improves performance on both the Rand. Set and the Prod. Set. However, excessive augmentation (at level~3) degrades performance and increases instability, especially on the Prod. Set. This reveals a trade-off between generalization, promoted by the randomness of augmentation, and stability, provided by the consistency of the deterministic DBA schedule. In our experiments, an intermediate augmentation level (level~2) achieves the best balance, outperforming all random baselines on the Prod. Set. This demonstrates the effectiveness of our approach, which combines a deterministic, semantically grounded keypose selection algorithm with controlled random augmentation to generalize effectively to real production scenarios.

%% file: figures/sota_comparison_table.tex
\begin{table*}[t]
\centering
% --- Further reduced space between columns ---
\setlength{\tabcolsep}{2.5pt}
% \caption{Comparison of our model against SOTA baselines across both our In-House Dataset (Keyframe-Based) and the LaFAN1 Dataset~\cite{harvey2020robust} (MOCAP).}
\caption{Comparison of our model against SOTA baselines across both our In-House Dataset (Keyframe-Based) and the LaFAN1 Dataset~\cite{harvey2020robust} (MOCAP). Our AIS-BiLSTM model outperforms the state-of-the-art baselines on all metrics for the Keyframe-Based dataset, and achieves competitive results on LaFAN1, highlighting its effectiveness for production-style animation as well as general motion data.}
\label{tab:sota_comparison}
\begin{tabular}{@{}lcccccccccccccccccc@{}}
\toprule
\multirow{3}{*}{Model} & \multicolumn{6}{c}{\textbf{Keyframe-Based Dataset}} & \multicolumn{12}{c}{\textbf{MOCAP Dataset}} \\
\cmidrule(lr){2-7} \cmidrule(lr){8-19}
& \multicolumn{2}{c}{\textbf{Algo. Set}} & \multicolumn{2}{c}{\textbf{Rand. Set}} & \multicolumn{2}{c}{\textbf{Prod. Set}} & \multicolumn{4}{c}{L2P $\downarrow$} & \multicolumn{4}{c}{L2Q $\downarrow$} & \multicolumn{4}{c}{NPSS $\downarrow$} \\
\cmidrule(lr){2-3} \cmidrule(lr){4-5} \cmidrule(lr){6-7} \cmidrule(lr){8-11} \cmidrule(lr){12-15} \cmidrule(lr){16-19}
& STL1$\downarrow$ & NPSS$\downarrow$ & STL1$\downarrow$ & NPSS$\downarrow$ & STL1$\downarrow$ & NPSS$\downarrow$ & 5 & 10 & 15 & 30 & 5 & 10 & 15 & 30 & 5 & 10 & 15 & 30 \\
\midrule
CondMDI & 17.42 & 3.65 & 15.09 & 2.27 & 33.06 & 5.09 & - & - & - & - & - & - & - & - & - & - & - & - \\
$\Delta$-interpolator & 4.85 & 1.02 & 5.87 & 1.09 & 3.51 & 0.59 & 1.65 & 2.90 & 4.16 & 8.52 & 0.94 & 1.27 & 1.57 & 2.65 & 0.43 & 0.62 & 1.10 & 3.39 \\
CITL & 12.61 & 1.75 & 12.15 & 1.47 & 27.47 & 2.49 & 1.61 & \textbf{2.37} & \textbf{3.39} & \textbf{7.47} & 1.01 & 1.21 & 1.39 & 1.88 & 0.43 & 0.66 & \textbf{0.88} & \textbf{2.76} \\

\textbf{AIS-BiLSTM (ours)} & \textbf{1.53} & \textbf{0.46} & \textbf{3.87} & \textbf{1.05} & \textbf{1.77} & \textbf{0.40} & \textbf{1.29} & 2.38 & 3.62 & 8.15 & \textbf{0.81} & \textbf{1.08} & \textbf{1.34} & \textbf{1.74} & \textbf{0.25} & \textbf{0.57} & 0.98 & 3.29 \\
\bottomrule
\end{tabular}
\end{table*}

%% file: sections/experiments/comparison_sota.tex
\subsection{Comparison with State-of-the-Art Methods}

We benchmark AIS-BiLSTM against methods that demonstrated state-of-the-art performance in sparse motion inpainting, i.e.\ motion in-betweening with sparse input keyframes---as opposed to motion completion with dense input context, which is the major setting in prior work. Specifically, we compare our method with the Transformer-based models $\Delta$-Interpolator~\cite{oreshkin2022delta} and CITL~\cite{mo2023citl}, and with the diffusion-based method CondMDI~\cite{cohan2024condmi}, which we adapted by removing its text-conditioning module.

\paragraph{In-House Animation Data}
All methods are retrained on our In-House dataset with identical data preprocessing and loss functions. Our AIS-BiLSTM is trained following the DBA schedule with the second Augmentation Level. SOTA methods are trained with their original input keypose schedules (a fixed 5-frame interval for $\Delta$-Interpolator, and random schedules for CITL and CondMDI). For a fair comparison, we performed hyperparameter tuning for each baseline using Optuna~\cite{akiba2019optuna}, exploring 50 trials per method (see Supplementary Material for details) to determine the optimal optimization and architectural parameters.

Results in Table~\ref{tab:sota_comparison}~(left) confirm the effectiveness of our contributions: AIS-BiLSTM consistently outperforms all baselines across test sets. Qualitatively (Fig.~\ref{fig:ctrl:sota}), \textbf{CondMDI} generates complex breakdown poses but is subject to jitter and temporal instability. \textbf{CITL} produces over-smoothed motions that drift from the input keyposes. The \textbf{$\Delta$-Interpolator} exhibits greater stability and better adherence to the input keyposes, which accounts for its higher quantitative scores. However, it tends to collapse toward near-zero offsets and produce overly smooth, SLERP-like trajectories, failing to reproduce the stylized, snappy variations present in the ground-truth motion. In contrast, our \textbf{AIS-BiLSTM} generates both stable and stylistically accurate motion that closely matches the ground-truth, effectively synthesizing complex breakdown poses and sharp transitions while adhering to input keyposes.

\paragraph{Motion Capture Data}
To assess the generalizability of our AIS-BiLSTM beyond our In-House dataset, we additionally train and evaluate our model on the LaFAN1 mocap benchmark~\cite{harvey2020robust} in a sparse input setting. For fairness, all models were integrated into the official CITL codebase, and our AIS-BiLSTM was trained with the same random input keypose schedule as prior work to match the evaluation protocol. As shown in Table~\ref{tab:sota_comparison}~(right), our model achieves state-of-the-art results for 5- and 10-frame gaps, while remaining competitive at longer horizons. This highlights its strength in short-term in-betweening---the dominant case in keyframe-based animation production---whereas CITL, based on a Transformer architecture in a direct synthesis mode, is better suited to very sparse, long-horizon predictions. Qualitative results in Fig.~\ref{fig:lafan1_renderings} further illustrate that our model is capable of producing smooth human motion with correct intermediate poses.

%% file: sections/experiments/production_workflow_evaluation.tex
\subsection{Production Workflow Evaluation}

As shown in Fig.~\ref{fig:renderings_sequences}, our model produces high-quality in-betweens that are superior to the Autodesk Maya spline baseline and that require only minor animator adjustments. To quantify the practical gains in production, we conducted a comparative study with two professional animators from an external studio, measuring our AI-assisted workflow against a fully manual process, on a total of 70.2 seconds of animation. Starting from a set of professionally authored block keyposes, animators completed the in-betweening phase under two conditions. In the \textbf{manual workflow}, they crafted the in-betweening motion manually from the block keyposes. In the \textbf{AI-assisted workflow}, the AIS-BiLSTM model predicted the in-betweening poses, and the animators manually performed the necessary retakes on the predictions to achieve production quality.

The results show a substantial acceleration: the manual workflow required \textbf{289 minutes}, whereas the AI-assisted workflow took only \textbf{83 minutes} for the animators. This corresponds to an approximate \textbf{3.5$\times$ speedup}, or a \textbf{71\% reduction} in human manual effort, demonstrating the potential of our method to significantly improve animation production efficiency.

%% file: sections/limitations_future_work.tex
\section{Limitations and Future Work}

Our method has three main limitations. \textbf{(1) Sensitivity to the input keypose schedule}: the model relies on the DBA as a proxy, leading to potential failures when production schedules differ from the training schedule (Fig.~\ref{fig:block_schedule_impact}). More accurate training schedules, obtained via algorithms that identify block poses in keyframe-based animation, or studio processes storing ground-truth block keyposes, could further improve performance. \textbf{(2) Limited long-horizon prediction}: our Bi-LSTM encoder is designed for the short gaps typical of production in-betweening and cannot infer missing content over large gaps (Fig.~\ref{fig:lafan1_renderings}). Combining the AIS layer with a Transformer-based encoder could address this. \textbf{(3) Character-specific training}: each character with its own rig requires a dedicated model. Fig.~\ref{fig:pato_sitting_jumping} shows a model trained on a different character capturing the overall style and typical poses, though effects like follow-through can fail. A unified multi-character model handling different rigs is a promising direction: trained on all character data, it could learn shared motion priors such as common style, similar breakdown poses, and standard animation effects like anticipation and follow-through.

%% file: sections/conclusion.tex
\section{Conclusion}

This work presents a motion in-betweening approach tailored to the specific characteristics of keyframe-based animation in production. Our method explicitly accounts for the \textit{nature of the data} through the Adaptive Interpolation–Synthesis (AIS) layer, which models each predicted pose as a mixture of directly synthesized breakdown keyposes and explicitly interpolated poses. The learned interpolation module enables the model to reproduce arbitrary \textit{motion styles} without bias toward smooth motion, while a deterministic input keypose schedule mitigates ambiguities introduced by snappy, discontinuous motion and improves stylistic consistency. Finally, our \textit{problem formulation} directly follows real production workflow---where in-betweening focuses on style rather than content generation---by selecting semantically meaningful keyposes as input and employing an LSTM encoder suited to short temporal gaps.

Overall, this work underscores the value of aligning deep learning systems with domain-specific creative workflows. By training exclusively on studio-owned data, supporting editability, and integrating seamlessly into production pipelines, our method provides a practical and ethical tool that enhances animator creativity, paving the way for more efficient and collaborative production pipelines.

%% file: figures/curves_detailed.tex
\newcommand{\triptych}[3]{%
  \begin{minipage}[c]{0.44\textwidth}
    \centering
    \includegraphics[width=\linewidth]{#1}%
  \end{minipage}\hspace{-0.5em}%
  \begin{minipage}[c]{0.14\textwidth}
    \centering
    \includegraphics[width=\linewidth]{#2}%
  \end{minipage}\hspace{-0.5em}%
  \begin{minipage}[c]{0.44\textwidth}
    \centering
    \includegraphics[width=\linewidth]{#3}%
  \end{minipage}%
}

\begin{figure*}[t]
  \centering

  % ========= 1) EXP: last_layer — direct_synthesis =========
  \begin{subfigure}{\textwidth}
    \triptych
      {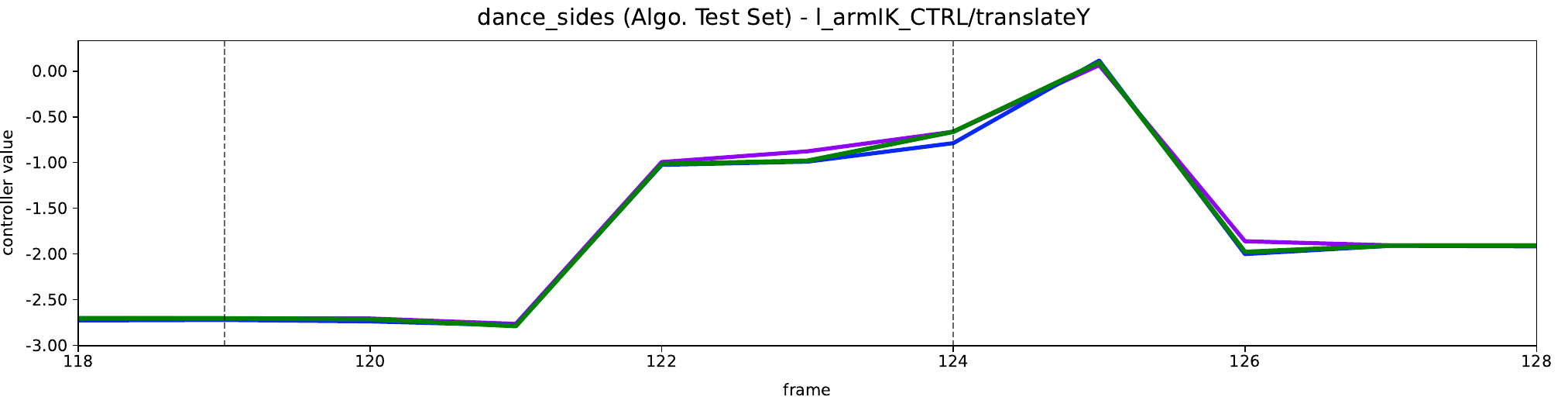}
      {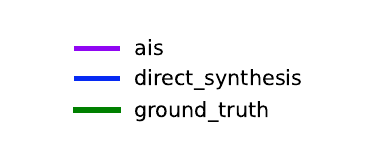}
      {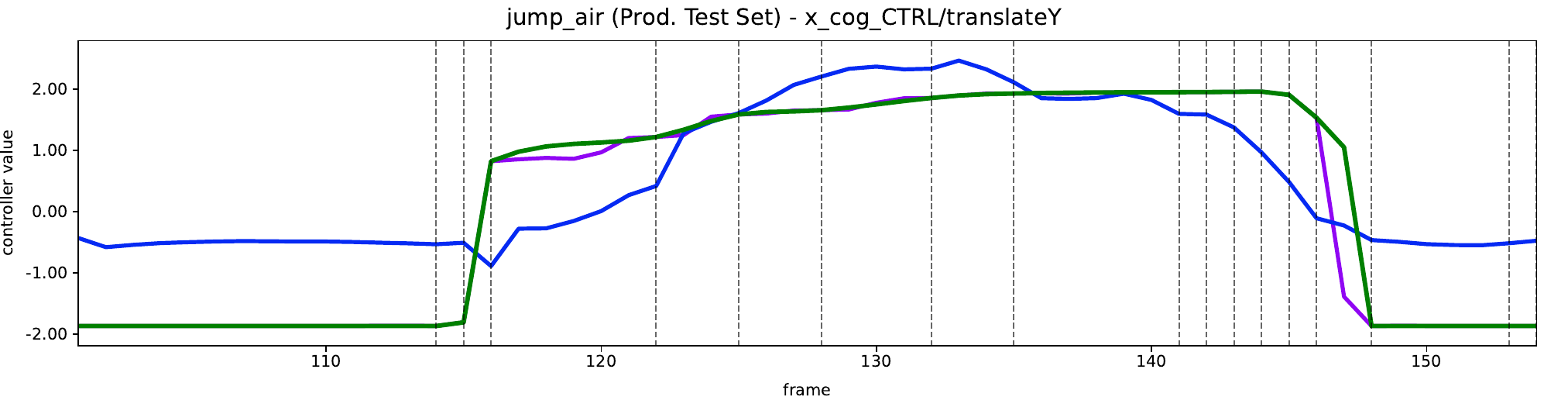}
    \subcaption{\textbf{Prediction Head Ablation Study: Direct Synthesis.} The direct synthesis head can generate complex motions (left) but tends to drift from the input values and to introduce noise on simpler trajectories (right).
      In contrast, our AIS Layer reproduces complex and simple motions with high fidelity to the ground truth.}
    \label{fig:ctrl:lastlayer:direct}
  \end{subfigure}

  % ========= 2) EXP: last_layer — ais mathematical interpolation =========
  \begin{subfigure}{\textwidth}
    \triptych
      {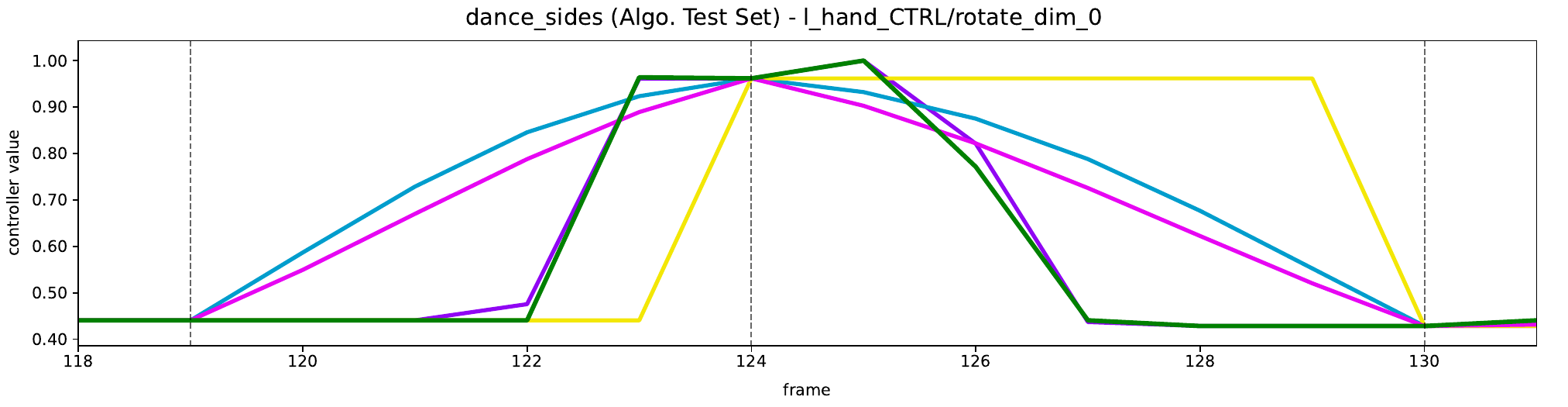}
      {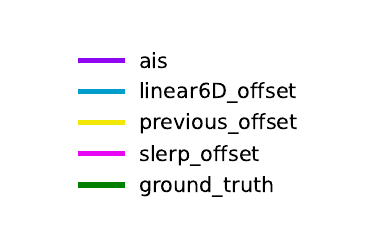}
      {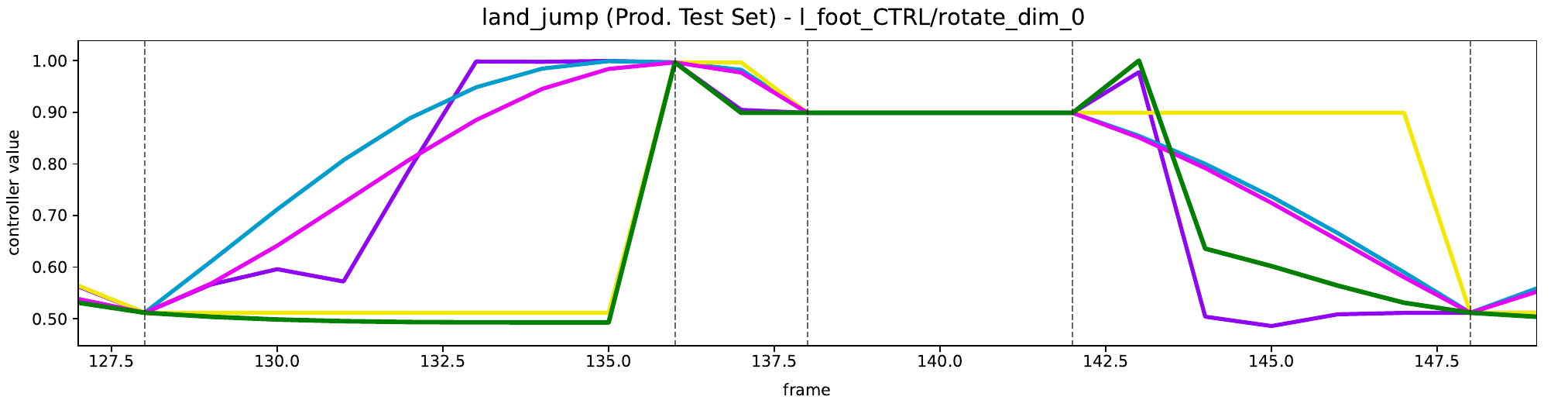}
    \subcaption{\textbf{Prediction Head Ablation Study: Fixed interpolation + learned offset.} Fixed interpolation heads (e.g., SLERP/linear-6D/previous) with a learned offset reproduce their fixed interpolation component, predicting a near-zero offset.}
    \label{fig:ctrl:lastlayer:mathinterp}
  \end{subfigure}

  % ========= 3) EXP: last_layer — ais learned interpolation =========
  \begin{subfigure}{\textwidth}
    \triptych
      {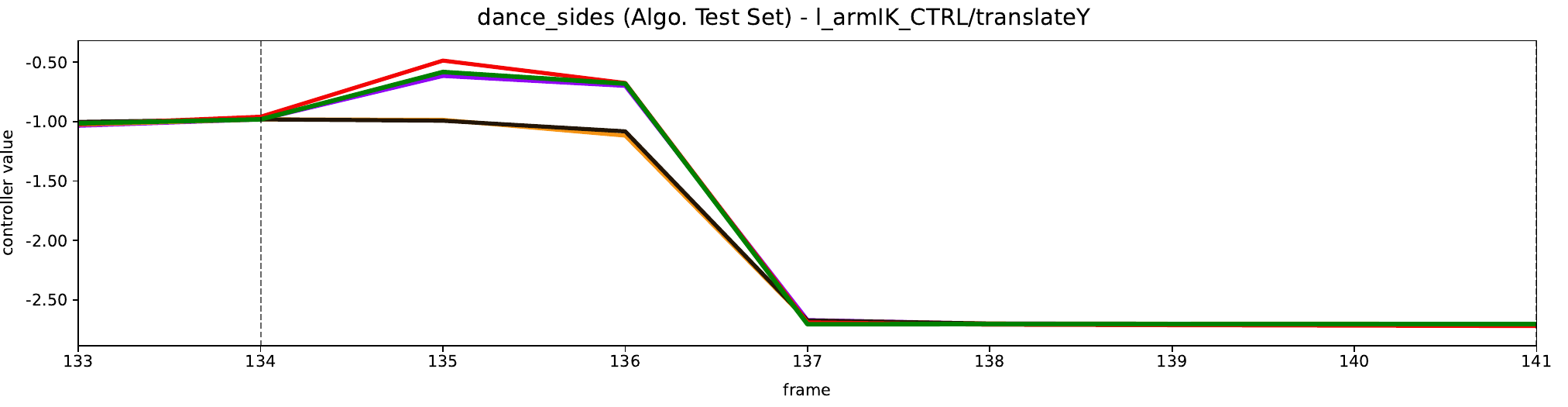}
      {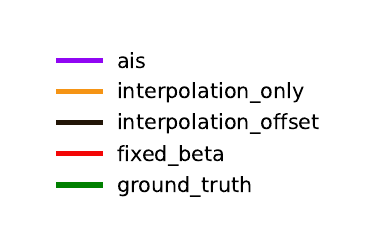}
      {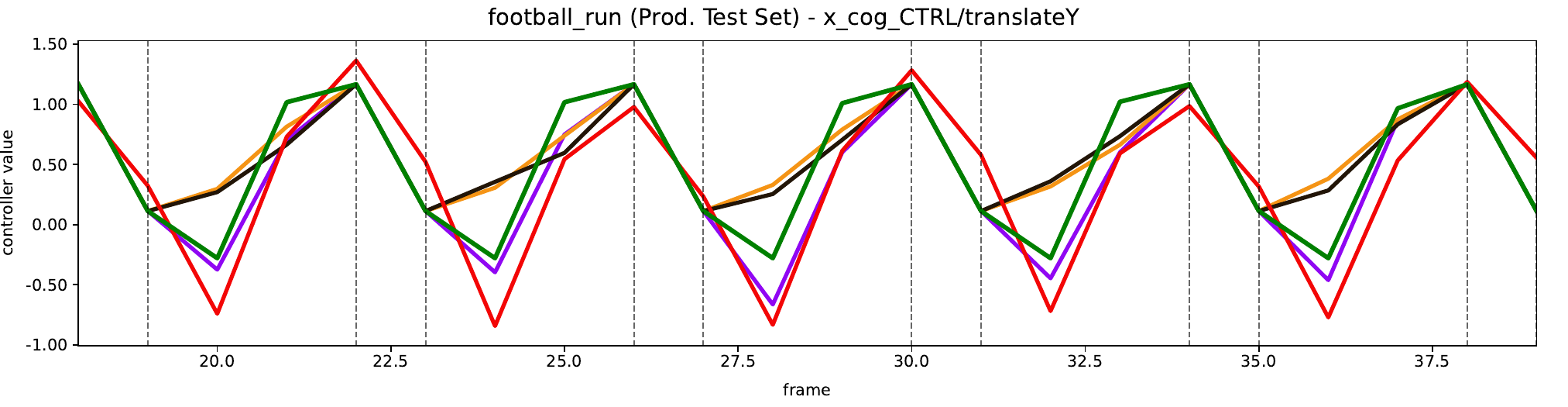}
    \subcaption{\textbf{Prediction Head Ablation Study: Learned Interpolation Only and with Offset.} Learned interpolation variants without a gating mechanism produce stable motions but fail at generating complex motions that go beyond the ranges of the input keyposes. In contrast, the AIS layer---through its gating mechanism---achieves both stability and expressivity across diverse motion amplitudes.}
    \label{fig:ctrl:lastlayer:learnedinterp}
  \end{subfigure}

  % ========= 4) EXP: block_schedule — random =========
  \begin{subfigure}{\textwidth}
    \triptych
      {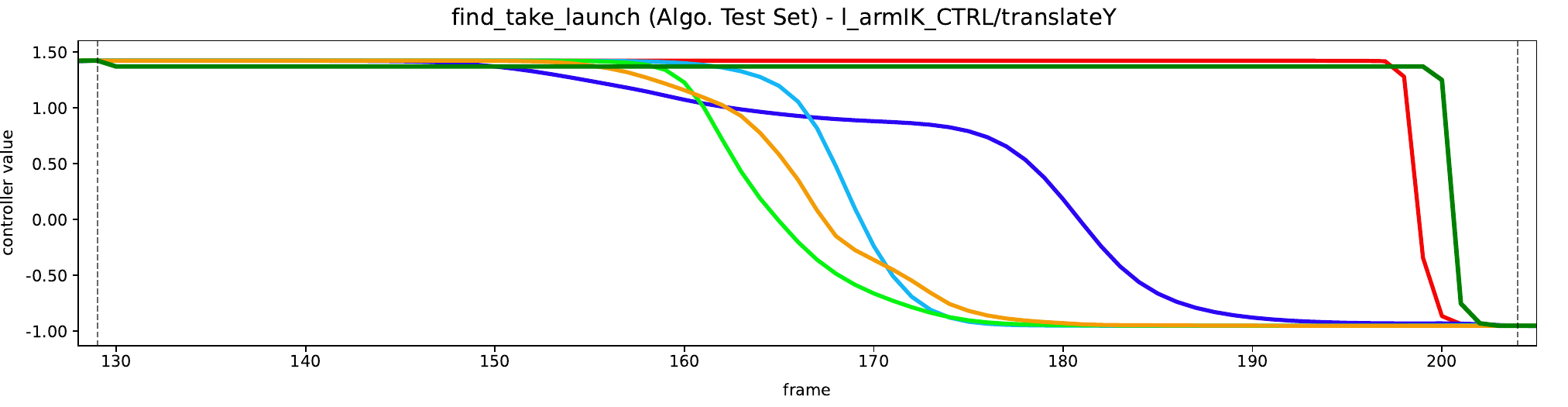}
      {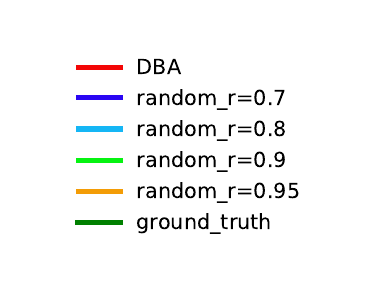}
      {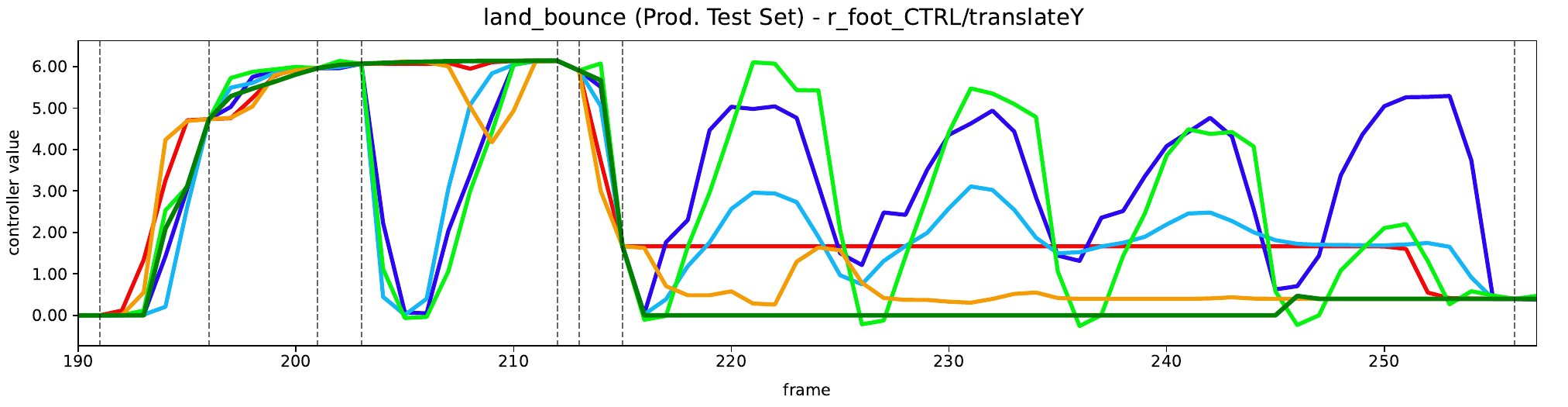}
    \subcaption{\textbf{Training Input Keypose Schedule Ablation Study: DBA vs Random Schedule.}
      Models trained with a random input keypose schedule fail to learn the snappy motion style, yielding overly averaged transitions (left) and unstable predictions with extra, undesirable motions (right).
      In contrast, the model trained with the DBA input keypose schedule produces more stable trajectories and better preserves the expressive motion style.}
    \label{fig:ctrl:blocksched:random}
  \end{subfigure}

  % ========= 5) EXP: block_schedule — augmented =========
  \begin{subfigure}{\textwidth}
    \triptych
      {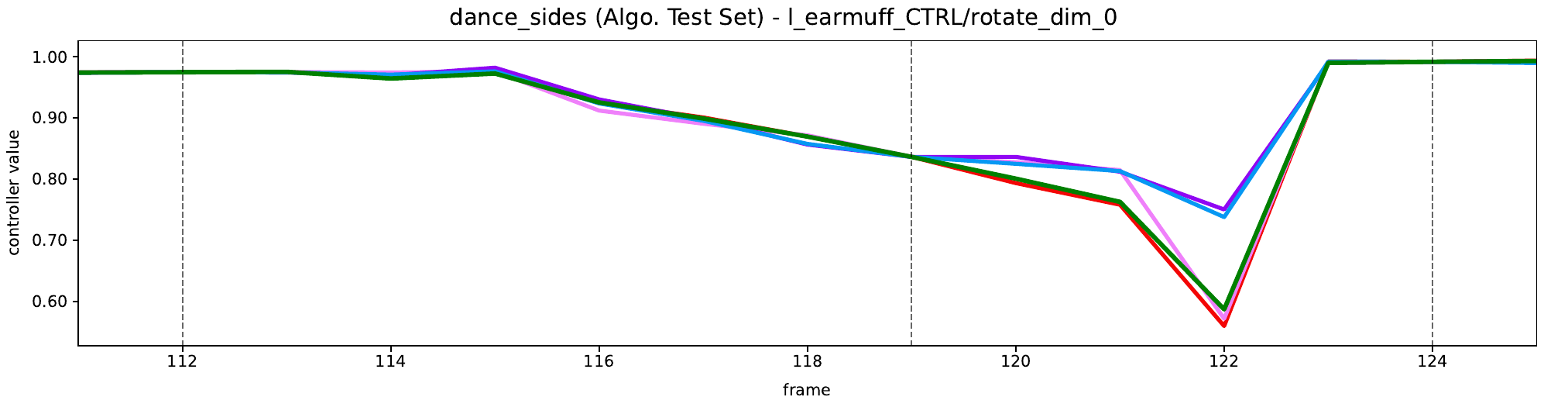}
      {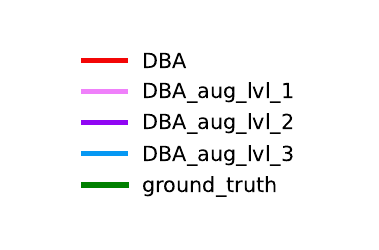}
      {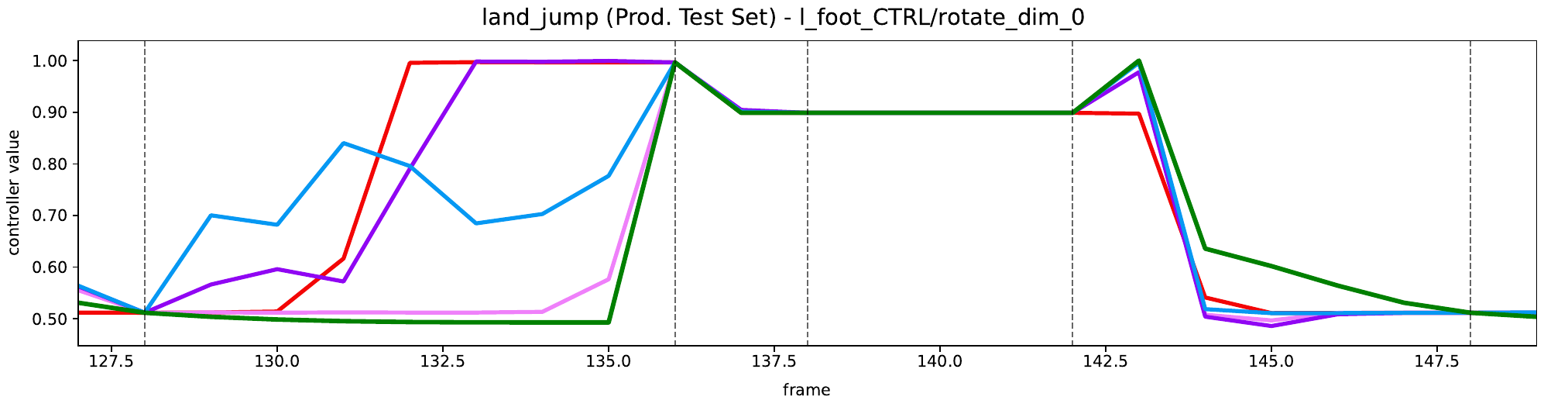}
    \subcaption{\textbf{Training Input Keypose Schedule Ablation Study: Augmentations on DBA.}
      Augmenting the DBA input keypose schedule slightly degrades performance on the DBA-based test set (left) but improves generalization on production data (right).
      However, excessive augmentation (e.g., level~3) introduces instability, revealing a trade-off between generalization and motion stability.}
    \label{fig:ctrl:blocksched:aug}
  \end{subfigure}

  % ========= 6) EXP: best_models =========
  \begin{subfigure}{\textwidth}
    \triptych
      {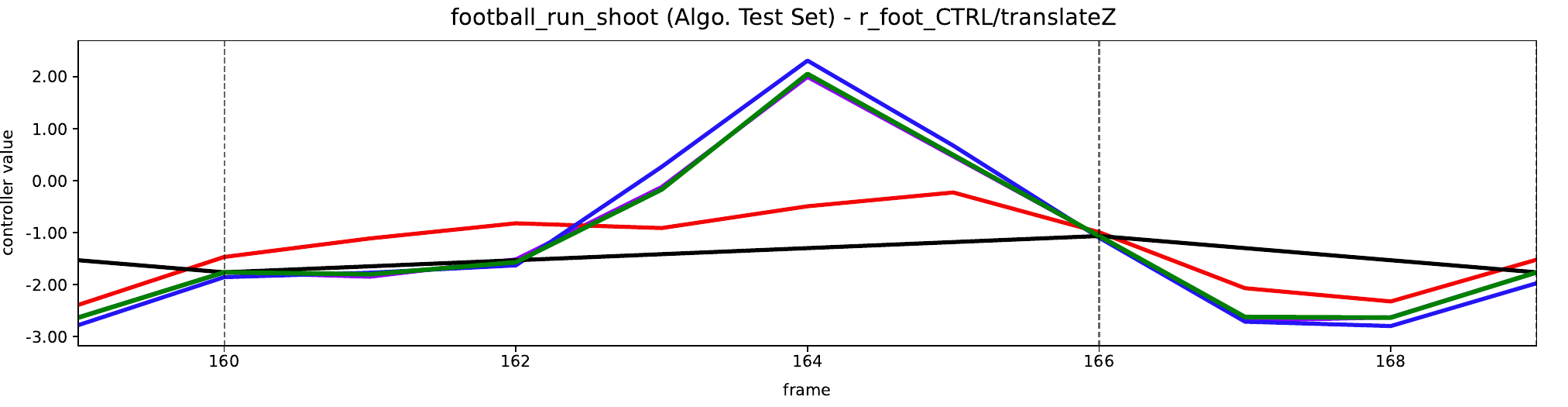}
      {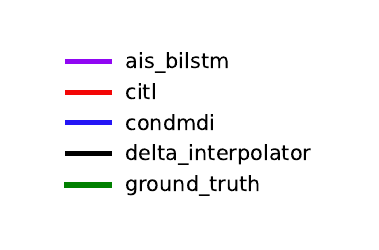}
      {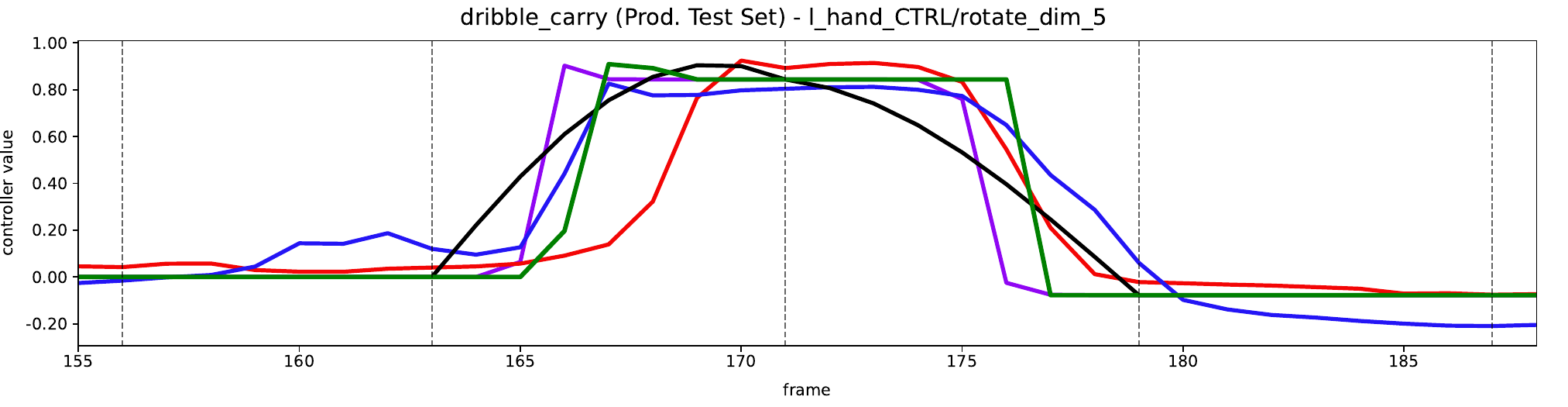}
    \subcaption{\textbf{State-of-the-Art Comparisons.}
      CondMDI and CITL demonstrate generative capabilities for typical motion cycles (left) and overshoot effects (right) but suffer from instability and input drift on simple motions.
      The $\Delta$-interpolator reproduces the SLERP interpolation almost exactly, yielding a near-zero offset---unsuitable for the snappy animation style.
      In contrast, our AIS-BiLSTM closely reproduces the ground-truth motion with both stability and fidelity.}
    \label{fig:ctrl:sota}
  \end{subfigure}

  \caption{\textbf{Qualitative Controller-Curve Comparisons.}
    We present qualitative results illustrating generated animation curves for various controller types (e.g., IK translations and components of the 6D rotations).
    For each study, we show results on the Algorithmic Test Set (left column) and the Production Test Set (right column).
    In all plots, the ground truth (green) is compared against the predictions of the different models, while vertical dotted lines indicate the input keyposes.}
  \label{fig:controllers:big}
\end{figure*}

%% file: figures/renderings_sequences.tex
\begin{figure*}[t!]
  \includegraphics[width=\linewidth]{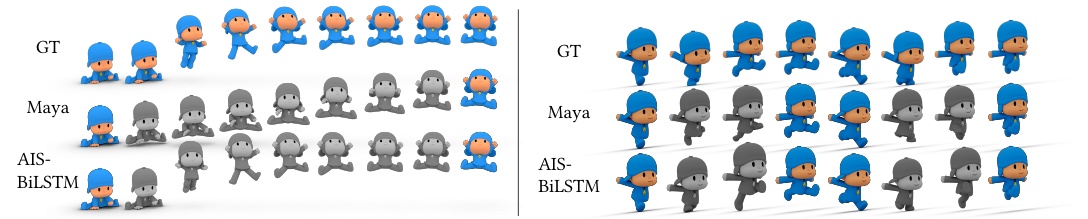}
  \captionsetup{skip=4pt}
  \caption{Qualitative results from the Algo. Test Set (left) and the Prod. Test Set (right); ground truth and input keyposes in color, predictions in grayscale. We compare ground truth (GT) and Maya's native spline interpolation (Maya) against our method (AIS-BiLSTM). Spline interpolation fails to create breakdown poses, produces self-penetration artifacts, and yields overly smooth motion, while our method generates plausible breakdowns, reproduces effects such as Ease In, and replicates the intended snappy style with held poses.}
  \label{fig:renderings_sequences}
\end{figure*}

%% file: figures/lafan1_renderings.tex
\begin{figure*}[t!]
  \centering
  \includegraphics[width=\textwidth]{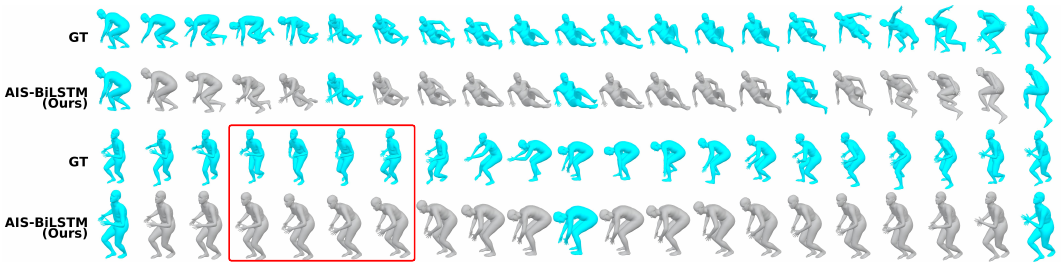}
  \captionsetup{skip=4pt}
  \caption{Qualitative results of our AIS-BiLSTM model trained on the LaFAN1 dataset (GT/inputs: color, predictions: grayscale). The model successfully learns smooth, human-plausible motion styles, with correct intermediate poses. However, as shown in the second example (\textcolor{red}{red box}), the model fails to predict an intermediate step in the running sequence. This indicates a limitation in inferring missing content over extended intervals.}
  \label{fig:lafan1_renderings}
\end{figure*}

% \begin{figure*}[t!]
%   \centering
%   \includegraphics[width=\textwidth]{figures/material_lafan1_renderings/sample_id_11_start_40_n_frames_21_interval_10.pdf}
%   \caption{Qualitative results on the LaFAN1 dataset.}
%   \label{fig:lafan1_renderings_2}
% \end{figure*}

%% file: figures/block_schedule_impact.tex
\begin{figure*}[t!]
    \includegraphics[width=\linewidth]{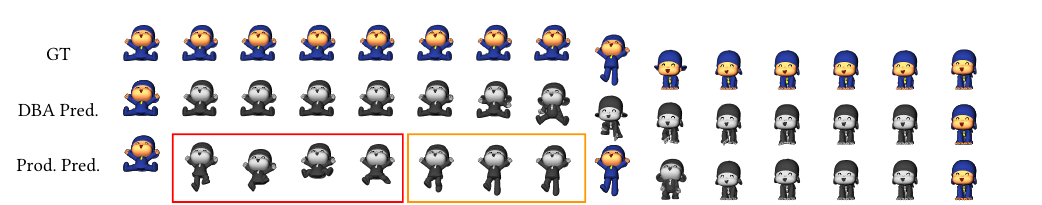}
    \captionsetup{skip=4pt}
    \caption{Illustration of sensitivity to the inference keypose schedule (GT/inputs: color, predictions: grayscale). "DBA Pred." infers from the DBA schedule used at training (two boundary keyposes), while "Prod. Pred." infers from a production schedule defined by animators, who added a middle keypose. The DBA prediction produces stable motion with held poses and a snappy transition, while the production prediction fails: the first pose is not held (\textcolor{red}{red box}) and the added middle keypose is over-held (\textcolor{orange}{orange box}).}
    \label{fig:block_schedule_impact}
\end{figure*}%

%% file: figures/pato_sitting_jumping.tex
\begin{figure*}[t!]
    \centering
    \includegraphics[width=\linewidth]{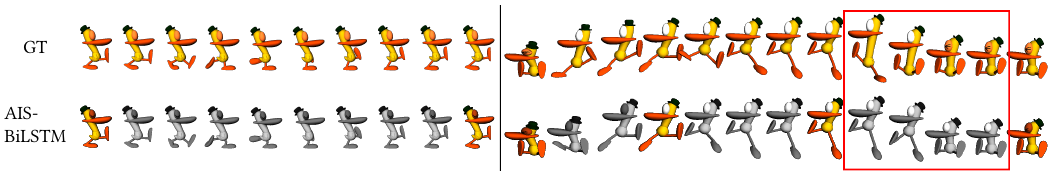}
    \captionsetup{skip=4pt}
    \caption{Qualitative results of our AIS-BiLSTM model trained on another character (GT/inputs: color, predictions: grayscale). (Left) Walking sequence: the model generates faithful breakdown keyposes of a typical walking cycle. (Right) Jumping sequence: the model respects the snappy transition style of the character; however, it fails to reproduce the follow-through effect seen in the ground-truth animation (\textcolor{red}{red}).}
    \label{fig:pato_sitting_jumping}
\end{figure*}%